\documentclass[%
 aip,
pof,
 amsmath,amssymb,
preprint,%
]{revtex4-1}

\usepackage{graphicx}
\usepackage{dcolumn}
\usepackage{bm}

\begin{document}


\title{Nonlinear periodic wavetrains in thin liquid films falling on a uniformly heated horizontal plate}

\author{Remi J. Noumana Issokolo}
\affiliation{Laboratory of Research on Advanced Materials and Nonlinear Sciences (LaRAMaNS),
Department of Physics, Faculty of Science, University of Buea P.O. Box 63, Buea, Cameroon.}
\author{Alain M. Dikand\'e}%
\email{dikande.alain@ubuea.cm} 
 \homepage{http://laramans.blogspot.com/.}
\affiliation{Laboratory of Research on Advanced Materials and Nonlinear Sciences (LaRAMaNS),
Department of Physics, Faculty of Science, University of Buea P.O. Box 63 Buea, Cameroon.}%

\date{\today}

\begin{abstract}
A thin liquid film falling on a uniformly heated horizontal plate spreads into fingering ripples that can display a complex dynamics ranging from continuous waves, nonlinear spatially localized periodic wave patterns (i.e. rivulet structures) to modulated nonlinear wavetrain structures. Some of these structures have been observed experimentally, however conditions under which they form are still not well understood. In this work we examine profiles of nonlinear wave patterns formed by a thin liquid film falling on a uniformly heated horizonal plate. In this purpose, the Benney model is considered assuming a uniform temperature distribution along the film propagation on the horizontal surface. It is shown that for strong surface tension but relatively small Biot number, spatially localized periodic-wave structures can be analytically obtained by solving the governing equation under appropriate conditions. In the regime of weak nonlinearity, a multiple-scale expansion combined with the reductive perturbation method leads to a complex Ginzburg-Landau equation, the solutions of which are modulated periodic pulse trains which amplitude, width and period are expressed in terms of characteristic parameters of the model.
\end{abstract}
\maketitle
\section{\label{sec:level1} Introduction}
A thin liquid film falling down a heated substrate spreads over the subsrate in form of fingering ripples exhibiting an extremely rich variety of spatio-temporal dynamics~\cite{kap1,kap2,frenk,chang,trif}. The system is an open-flow hydrodynamic problem which is convectively unstable~\cite{chang,greg,cras,kaba}, leading to a sequence of wave transitions beginning with the amplification of small-amplitude noises at the inlet, the filtering of linear stability and then a secondary modulational instability that transforms the primary wave field into a solitary-wave pulse~\cite{chang,thiele,orong}. This sequence of wave transitions is driven by the so-called long-wave instability mode, first observed experimentally by Kapitza and Kapitza~\cite{kap2}, and studied in detail by Benjamin~\cite{benj,benj1} who determined its threshold and established that a film falling down an inclined plane can only be destabilized for a Reynolds number larger than some critical value dependent on the inclination angle. The long-wave approximation was mathematically formulated for the first time by Benney~\cite{ben1} in terms of a single evolution equation for the thin-liquid free surface, later on solitary-wave solutions to the long-wave equation were numerically generated by several authors considering different ranges of characteristic parameters of this equation~\cite{kaba,thiele,orong,orong1,orong2,orong3,orong4,orong5,orong6,orong7,orong8,thiele1}. Thermally driven flows of free-surface thin-liquid films are ubiquitious in nature, in geophysics they occur in forms of gravity currents under water or lava flows~\cite{hup1,hup2}. In industry they are used in mass and heat transfers such as paintings, surface coatings and protections (e.g. TV screens, optical storage media, etc). \\
 Thin-liquid film flow on a heated horizontal wall can be described very simply as a shallow layer of fluid resting on a heated horizontal surface, which becomes unstable to both buoyancy-driven and thermocapillary convections~\cite{kap1,chang}. When the fluid layer is sufficiently thin, thermocapillary convection due to gradients in surface tension induces the Marangoni effect which turns to be the dominant source of instability in the film. It is most remarkable that due to the Marangoni effect the system can readily be regarded as one among fluid-mechanical problems that are described by Boussinesq's fourth-order nonlinear partial differential equations \cite{bouss,gala}. While the later fact has been recognized for some time now and considered in several past models for the thin-film flow, only by numerical simulations \cite{thiele,orong,orong1,orong2,orong3} it has been possible to gain insight onto the specific wave profiles driven by the competition between the Marangoni effect and the Rayleigh-Taylor instability. On this last point numerical simulations have revealed the existence of periodic wave patterns with permanent shapes characteristic of solitons and solitary waves. In refs. \cite{kaba,orong1,orong2,orong3,orong4} for instance, the existence and stability of these specific wave structures were discussed in both cases of film flowing on horizontal and slightly inclined plates. From the general standpoint of fundamental nonlinear theory the existence of solitary-wave structures requires a strong nonlinearity, but when the nonlinearity is weak either modulated envelopes with pulse or dark soliton profiles are to be expected. In the context of thin-liquid film flows the nonlinearity is introduced by Marangoni effects in the case of flow over an horizantal plate, and both by Marangoni effects and the Reynolds number when the plate is inclined. \\
 In this work we investigate conditions under which the Benney equation can lead either to well localized periodic wavetrain (i.e. rivulet-type) patterns, or to modulated nonlinear periodic wavtrain stuctures. To begin with, we shall introduce the model equation and show that in the specific case of a thin-liquid film falling on an horizontal plate, at small values of the Biot number a reasonable approximation enables analytical solutions describing a periodic pattern of nonlinear solitary waves reminiscent of rivulet patterns. Next, for the physical context where the nonlinearity is relatively small to totally balance the dispersion, a multiple-scale expansion combined with the reductive perturbation method \cite{frenk,kawa1} will lead to a cubic complex Ginzburg-Landau (CCGL) equation. Exact periodic solutions to this last equation will be obtained. 
 \section{\label{sec:one}
Mathematical Model and localized solitary-wave patterns}
\begin{figure} 
\includegraphics{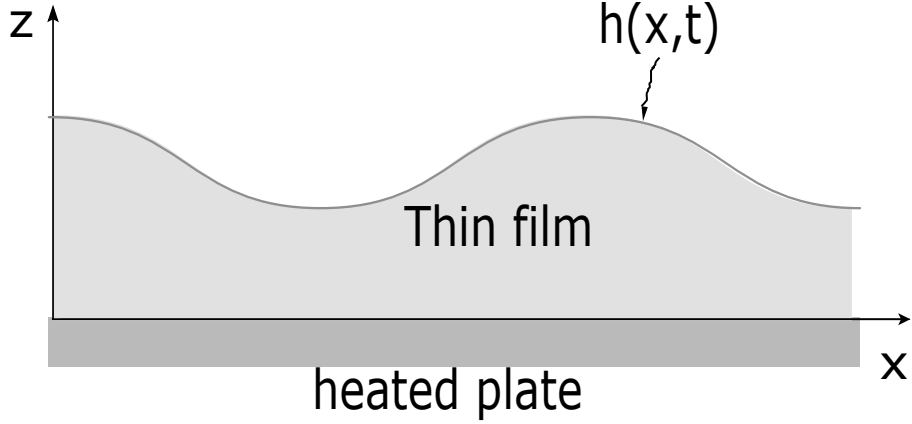}
\caption{\label{fig:one} (Color online) Geometry of a thin-liquid film flow along a uniformly heated horizontal plate.} 
\end{figure}

\begin{figure} 
\includegraphics{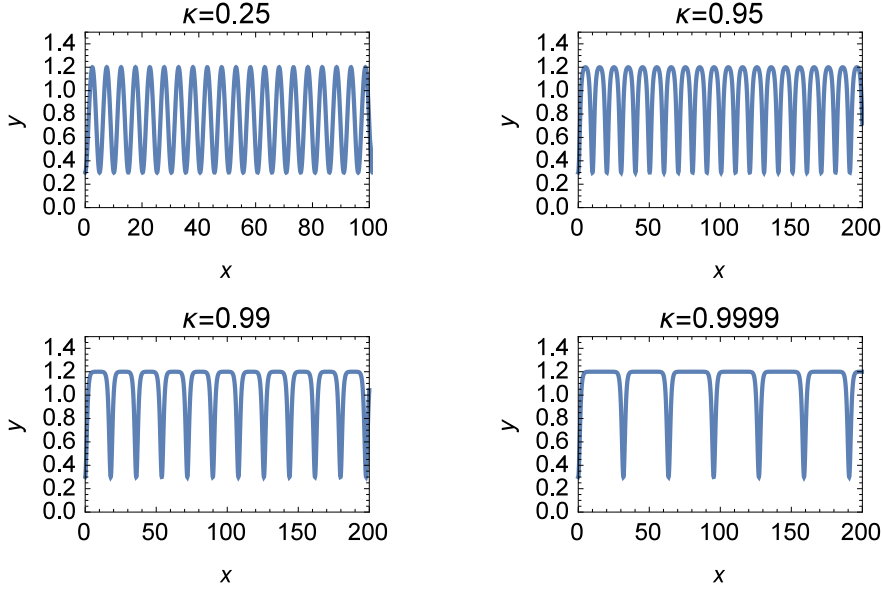}
\caption{\label{fig:two} (Color online) Sketch of the periodic wavetrain structure formula (\ref{e12a}), for four different values of the Jacobi elliptic modulus $\kappa$.} 
\end{figure}

The dominant picture in theoretical studies of the problem of thin-film flow over a uniformly heated plate is the long-wave instability proposed by Benney~\cite{ben1}, in which a small long-wave parameter is introduced in the hydrodynamic equation. Carrying out an expansion in this parameter ultimately leads to a single evolution equation, commonly referred to as the Benney equation \cite{frenk,ben1}. This picture, along with similar approaches, have proven very successful in determining the instability threshold while the evolution equations for the free surface, emerging from these techniques, have been applied to numerous problems ranging from Newtonian to non-Newtonian fluids. In the presence of surface tension, the propagation of the local spatial deformation of the film thickness $h(x,t)$ along the plate is governed by the nonlinear evolution equation~\cite{kaba,orong7}:
\begin{eqnarray}
 h_{t} = &-& [Rh^{3}+ \varepsilon(\frac{2}{15} R^{2}h^{6} h_{x} - C\frac{h^{3}}{3}h_{x} + S\frac{h^{3}}{3}h_{xxx} \nonumber \\
 &+& BM_{w}\frac{h^{2}}{2}\frac{(\delta + T_{w})h_{x}}{(1 + Bh)^{2}}  -  M_{w}\frac{h^{2}}{2} \frac{T_{w_{x}}}{1+Bh})]_x,
\label{e1}
 \end{eqnarray}
where $\beta$ is the angle of inclination of the plate from the horizontal surface and $T_w$ is the spatial temperature distribution along the plate. We are interested in a uniformly heated substrate so we assume $T_w$ to be a constant. Other parameters in the above equation are defined as:
\begin{eqnarray}
 S &=& \varepsilon^{2} \frac{\sigma_{\infty} h_{N}}{\rho \nu^{2}}, \hskip .02truecm M_{w} = \frac{Ma}{Pr},  \nonumber \\ 
 R &=& G \sin(\beta), \hskip 0.2truecm C = G \cos(\beta), \label{eda}
\end{eqnarray}
and represent respectively the surface tension ($S$), the effective Marangoni number ($M$), the Reynolds number ($R$) and the hydrostatic pressure ($C$). $\delta = \frac{\Delta \bar{T}}{\Delta \bar{T}_{w}}$ is the ratio of the characteristic temperature difference applied across the thin-liquid film layer ($\Delta \bar{T} = \bar{T}_{a} - \bar{T}_{\infty}$), to the average plate temperature $\bar{T}_{a}$, and $\bar{T}_{\infty}$ is the ambient temperature and the temperature difference along the plate is $\Delta \bar{T}_{w} = \bar{T}_{w_{max}} - \bar{T}_{w_{min}}$. \\ 
The first term in the right-hand side of eq. (\ref{e1}) is related to convective phenomena due to mean flow, the second term stands for inertia and is responsible for the hydrodynamic instability, the third term represents the stabilizing effect of the hydrostatic pressure, the fourth term accounts for surface tension and the fifth and sixth terms are associated with Marangoni effects \cite{kaba}. According to the last two terms in the right-hand-side of eq. (\ref{e1}), the thermocapillary effect in the film dynamics originates from two distinct sources namely: 
\begin{enumerate}
\item a perturbation of the plate temperature, induced by variations of the film thickness $h(x,t)$ when heat is transferred to the ambient medium ($B \neq 0$), 
\item a non-uniform heating of the wall (i.e. $T_{w}\neq$ constant). 
\end{enumerate}
In addition to fixing the plate temperature, we also focus on the film flow over an horizontal plate as illustrated in fig. \ref{fig:one}. 
For the flow geometry in fig. \ref{fig:one} the Reynolds number $R$ is zero, such that eq. (\ref{e1}) can be rewritten:
\begin{equation}
 h_{t} = - \frac{\partial}{\partial x} \left[\frac{\varepsilon h^{3}}{3}\frac{\partial}{\partial x} \left(Sh_{xx} + \frac{\partial P(h)}{\partial h}\right)\right],
 \label{e4}
\end{equation}
where:
\begin{equation}
 P(h) = C\frac{h^{2}}{2} - \frac{3 B M_{w} (\delta + T_{w})h}{2} Log\left(\frac{h}{1 + Bh}\right). \label{e5}
\end{equation}
Eq. (\ref{e4}) can also be expressed in the following general form:
\begin{equation}
 h_{t} = - \frac{\partial}{\partial x} \left[\varepsilon\frac{h^{3}}{3}\frac{\partial}{\partial x}\left(\frac{\partial F}{\partial h}\right)\right],
  \label{e6}
\end{equation}
characteristic of a variational problem where $F(h)$, the "energy functional", is defined as: 
\begin{equation}
 F(h) = \frac{1}{L} \int{\left[\frac{S}{2}h_{x}^{2} + P(h)\right] dx}. \label{e7}
\end{equation}
It is remarkable that $F(h)$ is similar to the energy functional of a system undergoing a Landau-type structural phase transition \cite{cow}, where $h(x,t)$ holds the role of the space-time varying order parameter while the first term in the integral accounts for the contribution from spatial distorsion of the order parameter. $P(h)$ in this case is the free energy associated with the homogeneous phase of the system. \\
The analogy with the Landau-type second-order phase transition, enables us determine the equilibrium spatial structure of the order parameter $h(x)$ by seeking for stationary solutions to the variational equation:
\begin{equation}
  h_x = \sqrt{-\frac{2}{S}\lbrack P(h) - P_0\rbrack},
   \label{e8}
\end{equation}
where $P_0$ is an integration constant. Solving eq. (\ref{e8}), it is useful to start by stressing that the solution of interest is the one describing profiles of the wave pattern forming on the film surface. For such solution, we can express the total film thickness $h(x)$ as the sum of a uniform (i.e. undeformed) thickness $h_0$, and small ripples of amplitudes $Y(x)$ representing spatial deformations of the film surface. That is we set $h = h_0(1 + y)$ (with $y=Y/h_0$), and assuming $Y<<h_0$ (i.e. $y<<1$) we can expand the free energy $P(h)$ in $y(x)$.
In the regime where the Biot number $B$ is small we can define a "small" expansion parameter:
\begin{equation}
\alpha  =  \frac{B h_{0}}{1 + B h_0}, \label{alph}
\end{equation}
in terms of which an expansion of $P(h)$ to the third order transforms eq. (\ref{e8}) into the first-order elliptic problem:
\begin{equation}
 h_z = \sqrt{A_3\,y^3 + A_2\,y^2 + A_1\,y + A_0}, \hskip 0.25truecm z= x\,\sqrt{\frac{2}{Sh^2_0}},
   \label{e8a}
\end{equation}
with: 
\begin{eqnarray}
 A_{3} & = &  -\frac{3 B M_w (\delta + T_w) h_0}{4}\lbrack (1 - \alpha^2) - \frac{2}{3}(1-\alpha^3)\rbrack, \nonumber \\
 A_{2} & = & -\frac{1}{2}C h_0^2 + \frac{3 B M_{w} (\delta + T_{w}) h_0}{4} (1 - \alpha)^2 , \nonumber \\
 A_{1} & = & -C h_0^2 + \frac{3 B M_{w} (\delta + T_{w}) h_{0}}{2} \left[ (1 - \alpha) + log \left(\frac{\alpha}{B}\right)\right], \nonumber \\
 A_{0} & = & -\frac{1}{2}C h_0^2 + P_0+\frac{3 \varepsilon B M_{w} (\delta + T_{w}) h_{0}}{2} log \left(\frac{\alpha}{B}\right). \label{coefs1}
\end{eqnarray}
Integration of eq. (\ref{e8a}) defines Weierstrass's elliptic functions \cite{abra}, which lead to the following nonlinear periodic wavetrain solution: 
\begin{eqnarray}
 y(x)&=& e_2 + (e_1-e_2)\,cn^2\left(x/\nu, \kappa \right), \label{e12a} \\
 \nu&=& \sqrt{\frac{2h_0^2S}{-(e_1-e_3)A_3}}, \hskip 0.3truecm \kappa =\sqrt{\frac{e_1-e_2}{e_1-e_3}}. \label{e12}
\end{eqnarray}
In the last solution $e_{i=1,2,3}$ are real roots of the cubic polynomial $F(y)= -2(y-e_1)(y-e_2)(y-e_3)$, generated by setting $P(y)=-A_3 F(y)$ which gives the following three transcendental equations: 
\begin{eqnarray}
e_1 + e_2 + e_3&=-&A_2/A_3, \nonumber \\
e_1e_2 + e_1 e_3 + e_2e_3&=& A_1/A_3, \nonumber \\
e_1 e_2 e_3&=& -A_0/A_3. \label{coefcub}
\end{eqnarray}
The quantity $cn$ in (\ref{e12a}) is a Jacobi elliptic function \cite{abra,dik1,dik2,dik3}, its modulus $\kappa$ is given in (\ref{e12}) in terms of $e_1$, $e_2$ and $e_3$ which, according to (\ref{coefcub}), are functions of physical parameters of the model. Hence the modulus $\kappa$ of the Jacobi elliptic function also is a function of physical parameters of the model. Though this dependence seems complicated when considering the relations between $e_{1,2,3}$ and parameters in the elliptic nonlinear equation (\ref{e8a}), more simple expressions of $\kappa$ as a function of physical parameters of the model are obtained for its two limiting values. Indeed by definition\cite{abra} the modulus $\kappa$ must obey $0\leq k\leq 1$, for the first characteristic value i.e. $\kappa \rightarrow 0$ the expression of $\kappa$ in formula (\ref{e12}) suggests that $e_1=e_2$. Therefore, from the set eqs. (\ref{coefcub}) we can easily derive relations between parameters of the model for which $cn() \rightarrow \cos()$.  On the other hand, the upper value $\kappa= 1$ implies $e_2=e_3$, which also provides a relationship between the model parameters for which the Jacobi elliptic function $cn()\rightarrow sech()$. \\
The Jacobi elliptic function $cn$ is periodic with respect to $x$, meaning that the solution (\ref{e12a}) can be regarded as a periodic wavetrain of spatially localized rivulets forming from the thin-liquid film spreading along the horizontal plate. In fig. \ref{fig:two} we sketch the solution (\ref{e12a}) for four distinct values of $\kappa$, corresponding to four distinct relationships between physical parameters of the model i.e. $\kappa=0.25$, $0.95$, $0.99$ and $0.9999$. 
The figure shows that for small values of $\kappa$, the relative deformations $y$ of the thin-liquid film are periodic ripples with harmonic profiles. As we increase $\kappa$ ($\kappa=0.95$) the ripples gain in anharmonicity, a feature noticeable in their typical fingering shapes \cite{legros} reminiscent of rivulet structures \cite{legros1,legros2,legros3}. As $\kappa\rightarrow 1$ the ripples are more and more broad (graphs for $\kappa=0.98$ and $\kappa=0.99$). It is worth noting that $\nu$ in the expression of $y(x)$ given by formula (\ref{e12a}), is the average width (i.e. the width at half crest) of individual ripples in the periodic wavetrain. According to formula (\ref{e12}), an increase of the surface tension will broaden the ripples. The most telltale physical implication of this broadening should be looked in the change of the period of the wavetrain, which is given by the period of (\ref{e12a}) i.e. $T_p=2K(\kappa)\nu$. According to the analytical expression of $T_p$, an increase in the surface tension should cause an increase of the separation between ripples in the periodic wavetrain. In fact, this is expected to favor the formation of single solitary-wave structures similar to some of those obtained in recent numerical studies (see e.g. ref. \cite{orong2}).
\section{\label{sec:level3}Modulated nonlinear periodic wavetrains}
\begin{figure} 
\includegraphics{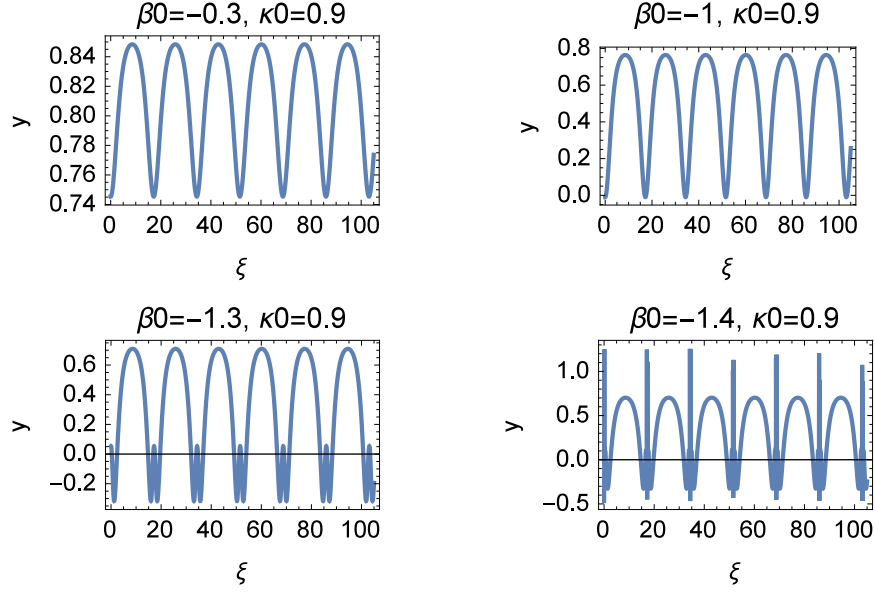}
\caption{\label{fig:three} (Color online) Profiles of the modulated-wave solution $y(\xi)$ given by (\ref{finsol}), for $\kappa_0=0.9$ and four different negative values of $\beta_0$ indicated in the graphs.} 
\end{figure}

\begin{figure} 
\includegraphics{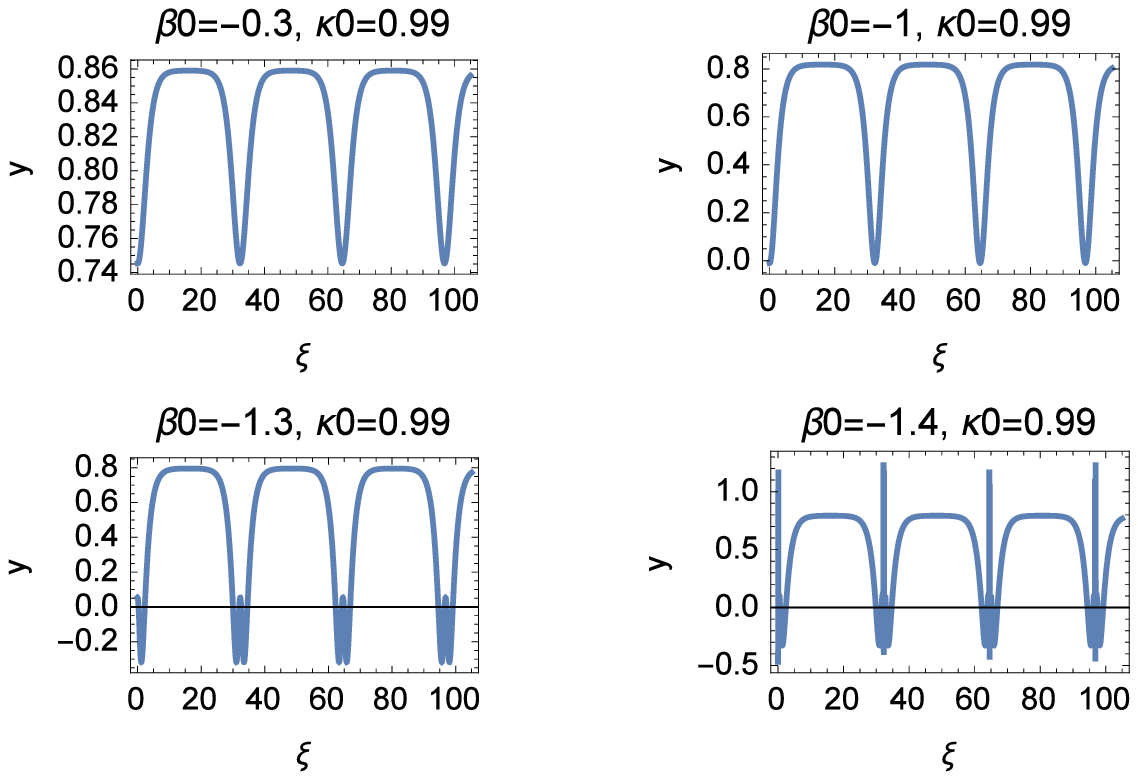}
\caption{\label{fig:four} (Color online) Profiles of the modulated-wave solution $y(\xi)$ given by (\ref{finsol}), for $\kappa_0=0.99$ and four different negative values of $\beta_0$ indicated in the graphs.} 
\end{figure}

\begin{figure} 
\includegraphics{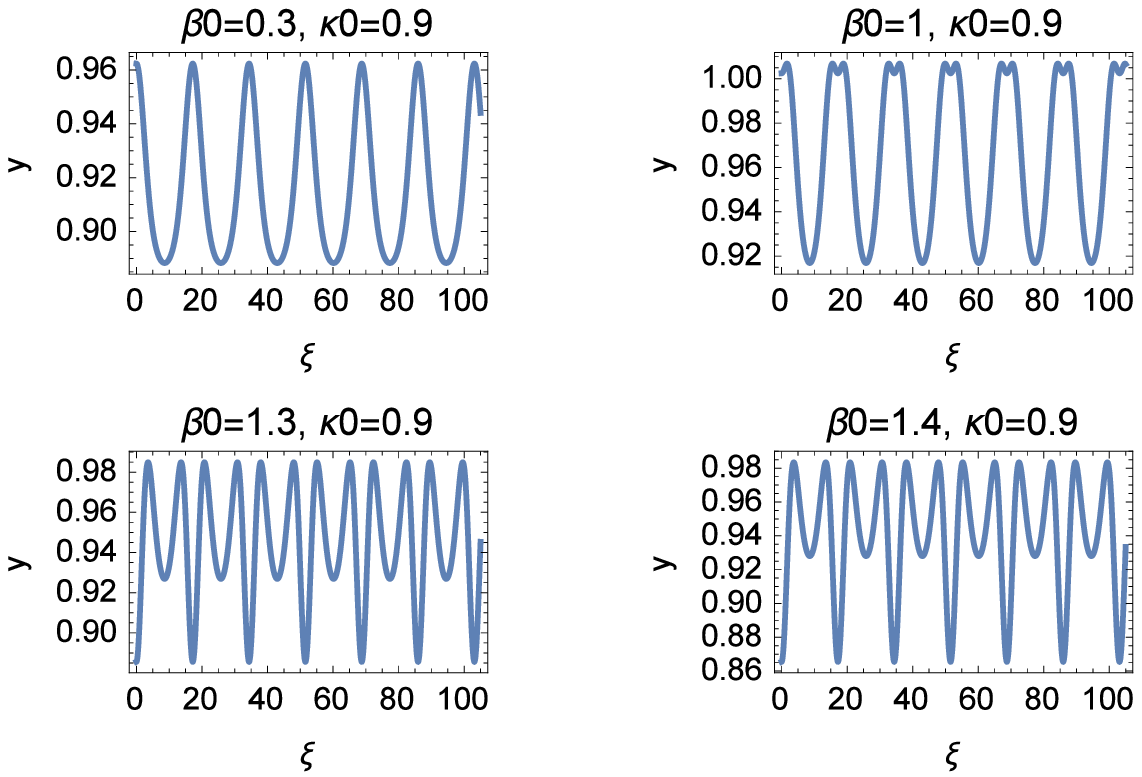}
\caption{\label{fig:five} (Color online) Profiles of the modulated-wave solution $y(\xi)$ given by (\ref{finsol}), for $\kappa_0=0.9$ and four different positive values of $\beta_0$ indicated in the graphs.} 
\end{figure}

\begin{figure} 
\includegraphics{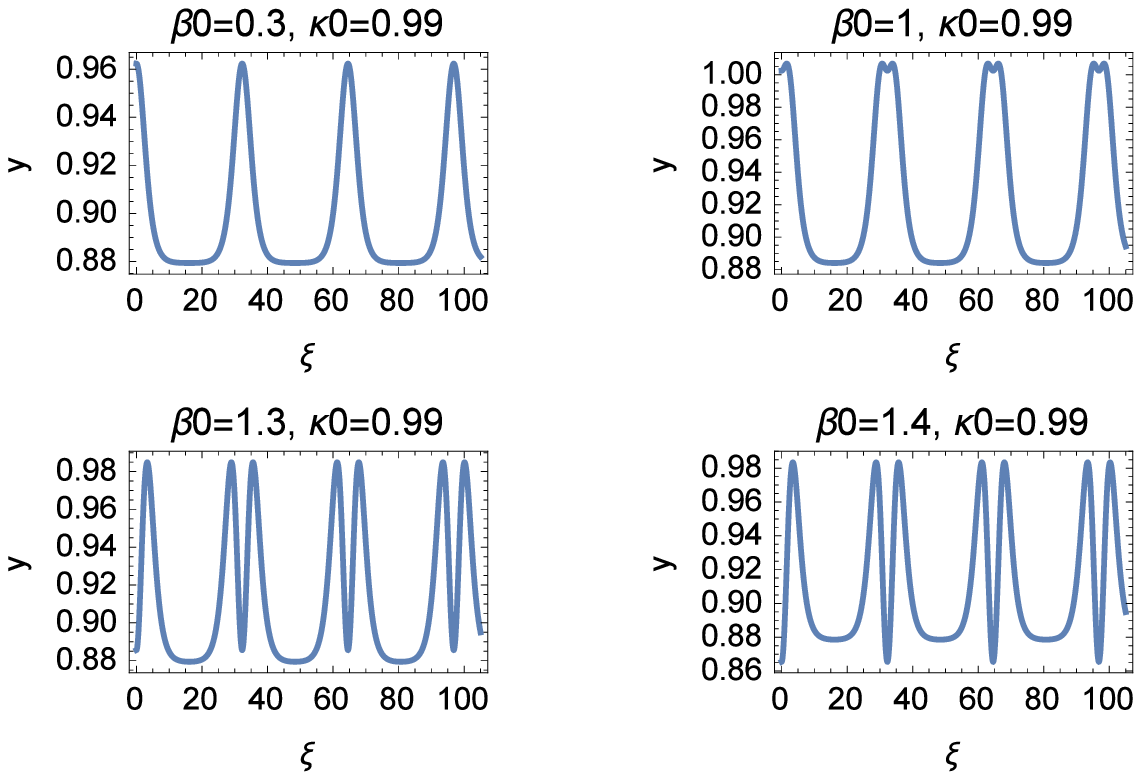}
\caption{\label{fig:six} (Color online) Profiles of the modulated-wave solution $y(\xi)$ given by (\ref{finsol}), for $\kappa_0=0.99$ and for different positive values of $\beta_0$ indicated in the graphs.} 
\end{figure}

When the dispersion becomes dominant over nonlinearity, the periodic spatial wavetrain generated by deformations of the thin-liquid film surface can develop spatio-temporal modulations \cite{kar,tap,jef,kim,zaka}. To investigate these weakly nonlinear periodic modulated-wave structures for the evolution equation (\ref{e1}), we shall apply a multiple-scale expansion combined with the reductive-perturbation method \cite{kawa1} on this equation. \\ 
To start let us express eq. (\ref{e1}) in terms of the small relative deformation $y$:
\begin{eqnarray}
 y_t &+& R h_{0}^{2}(1+y)^{2} y_{x} +  \varepsilon [\frac{2}{15} R^{2}h_{0}^{6}(1+y)^{6} y_{x} \nonumber \\
 &-& \frac{C h_{0}^{3}(1+y)^{3}}{3}y_{x}  + \frac{S h_{0}^{3}(1+y)^{3}}{3}y_{xxx} + \frac{\mu \alpha_{0}^{2}(1+y)^{2} y_{x}}{(1 + \alpha y)^{2}}]_{x} = 0.
\label{a1}
\end{eqnarray}
For small $y$, we can expand terms of the form $(1+y)^n$ in the last equation to the order $y^2$. Doing this eq. (\ref{a1}) reduces to:
\begin{eqnarray}
 y_t &+& a_1 y_{x} + b_1 y_{xx} + c_1 y_{xxxx} + a_2 y y_{x} \nonumber \\
&+& a_1 (y^{2}_{x} + y y_{xx}) + c_2 (y_x y_{xxx} + y y_{xxxx}) + a_1 y^2 y_x \nonumber \\
 &+& b_3 ( y^2 y_{xx} + 2 y y_{x}^{2}) + c_1(y^{2} y_{xxxx} + 2 y y_{x} y_{xxx}) = 0, 
 \label{e14}
\end{eqnarray}
where:
\begin{eqnarray}
a_1 &=& R h_{0}^{2}, \hskip 0.25truecm a_2 = 2 a_1, \hskip 0.25truecm c_1 = \frac{C h_{0}^3}{3a},  \hskip 0.25truecm c_2 = 3c_1, \nonumber \\
b_1 &=& \frac{2}{15} R^{2}h_{0}^{6} -\frac{C h_{0}^{3}}{3} + \mu \alpha_0^{2}, \hskip 0.25truecm \alpha_0^{2} = \frac{h_{0}^{2}}{(1 + Bh_{0})^{2}} \nonumber \\
b_2 &=& \frac{4}{5} R^{2}h_{0}^{6} - C h_{0}^{3} + 2\mu \alpha_{0}^{2}(1-\alpha), \nonumber   \\
b_3 &=& 2R^{2}h_{0}^{6} - C h_{0}^{3} + \mu \alpha_{0}^{2}(1 - 2\alpha + 3\alpha^{2}). \label{e12ac} 
\end{eqnarray}
The linearized part of (\ref{e14}) admits travelling-wave solutions of the form:
\begin{equation}
 y(x,t) = \phi\,exp \lbrace i(k x - wt)\rbrace +  c.c,
 \label{e15}
\end{equation}
where $\phi$ is the wave amplitude with c.c referring to its complex conjugate. $k$ is the wave number and $\omega$ is the wave frequency. Substituting  eq. (\ref{e15}) in the linearized version of eq. (\ref{e14}), we find the linear-wave dispersion relation:
\begin{equation}
 D(\omega,k) \equiv - i\omega + ia_1k - b_1k^{2} + c_1k^{4} = 0.
 \label{e16}
\end{equation}
As formula (\ref{e16}) suggests, for real values of the wave number $k$ the frequency $\omega$ is complex. Hence $\omega$ can be decomposed as $\omega=\omega_r+i\omega_i$, with:
\begin{equation}
 \omega_{r} = a_1k, \hskip 0.25truecm  \omega_i = b_1 k^{2} - c_1 k^4.
 \label{e17}
\end{equation}
We define the phase velocity $V_{p}$ as: 
\begin{equation}
 V_{p} = \frac{\omega_{r}}{k} = a_1 = R h_{0}^{2}.
 \label{e18}
\end{equation}
As we are interested in small-amplitude waves, we can readily consider the nonlinear evolution of an unstable wave 
in the regime where damping is not sensitive to the order $\zeta^2$, i.e. $\omega_{i} \sim \textit{O}(\zeta^{2})$ where $\zeta$ is a small perturbation. In this respect we introduce the new variables:
\begin{equation}
 x_{1} = \zeta x, \hskip 1.0truecm x_{2} = \zeta^{2} x, \hskip 0.25truecm t_{1} = \zeta t, \hskip 0.25truecm t_{2} = \zeta^{2} t, \hskip 0.25truecm ...,
 \label{e19}
\end{equation}
where $t$ and $x$ are fast-scale variables, while $x_{1}$ and $t_{1}$ and so on are slow-scale variables. Since these variables are assumed mutually independent, we can define temporal and spatial derivatives as:
\begin{equation}
\partial_{t}\rightarrow \partial_{t_{0}} + \zeta \partial_{t_{1}} + \zeta^{2}\partial_{t_{2}} + ...,  \hskip 0.25truecm \partial_{x}\rightarrow \partial_{x_{0}} + \zeta \partial_{x_{1}} + \zeta^{2}\partial_{x_{2}} + ... .
\label{e20}
\end{equation}
Inserting (\ref{e19}) and (\ref{e20}) in eq. (\ref{e14}), and expanding the solution $y$ i.e.:
\begin{equation}
 y(x, t)=\sum_{i=1}{\zeta^i y_i(x_i, t_i)}, \label{sumap}
\end{equation}
 we can re-express (\ref{e14}) formally as:
\begin{equation}
 L \left(\frac{\partial}{\partial t}, \frac{\partial}{\partial x}\right) y = - \zeta^{2} N_{2} - \zeta^{3}N_{3} ,
 \label{e21}
\end{equation}
where $L$ is a linear operator and $ N_{2}$ and $N_{3}$ are the nonlinear terms. More explicitely eq. (\ref{e21}) is of the form:
\begin{equation}
 (L_{0} + \zeta L_{1} + \zeta^{2}L_{2}+...)(\zeta y_{1} + \zeta^{2} y_{2} + \zeta^{3} y_{3}+...) = - \zeta^{2} N_{2} - \zeta^{3}N_{3} - ... ,
 \label{e22}
\end{equation}
where $L_{0}$, $L_{1}$ and $L_{2}$ are all linear operators of different orders in $\zeta$ and are given by: 
\begin{eqnarray}
 L_{0} &\equiv& \frac{\partial}{\partial t_{0}} + a_1 \frac{\partial}{\partial x_{0}} + b_1\frac{\partial^{2}}{\partial x_{0}^{2}} + c_1 \frac{\partial^{4}}{\partial x_{0}^{4}} ,\\
 L_{1} &\equiv& \frac{\partial}{\partial t_{1}} + a_1\frac{\partial}{\partial x_{1}} + 2b_1\frac{\partial^{2}}{\partial x_{0} \partial x_{1}} + 4c_1\frac{\partial^{4}}{\partial x_{0}^{3} \partial x_{1}}, \\
 L_{2} &\equiv& \frac{\partial}{\partial t_{2}} + a_1\frac{\partial}{\partial x_{2}} + b_1 \left(\frac{\partial^{2}}{\partial x^{2}_{1}} +2\frac{\partial^{2}}{\partial x_{0} \partial x_{2}}\right) + c_1\left(6\frac{\partial^{4}}{\partial x_{0}^{2} \partial x_{1}^{2}} + 4\frac{\partial^{4}}{\partial x_{0}^{3} \partial x_{2}}\right), \label{lin}
\end{eqnarray}
while the nonlinear terms $N_1$ and $N_2$ are given by:
\begin{eqnarray}
N_{1} &=& a_2 y_{1}\frac{\partial y_{1}}{\partial x_{0}} + b_2\left[y_{1}\frac{\partial^{2} y_{1}}{\partial x_{0}^{2}} + \left(\frac{\partial y_{1}}{\partial x_{0}}\right)^{2}\right] + c_2\left[y_{1} \frac{\partial^{4}y_{1}}{\partial x_{0}^{4}} + \frac{\partial y_{1}}{\partial x_{0}} \frac{\partial^{3} y_{1}}{\partial x_{0}^{3}}\right], \\ 
N_{2} &=& a_2\left[y_{1}\left(\frac{\partial y_{2}}{\partial x_{0}} + \frac{\partial y_{1}}{\partial x_{1}}\right) + y_{2}\frac{\partial y_{1}}{\partial x_{0}}\right] \nonumber \\
 &+& b_2\left[y_{1}\left(\frac{\partial^{2} y_{2}}{\partial x_{0}^{2}} + 2\frac{\partial^{2} y_{1}}{\partial x_{0} \partial x_{1}}\right) + y_{2}\frac{\partial^{2} y_{1}}{\partial x_{0}^{2}} + 2\frac{\partial y_{1}}{\partial x_{0}}\left(\frac{\partial y_{2}}{\partial x_{0}} + \frac{\partial y_{1}}{\partial x_{1}}\right)\right] \nonumber \\
&+& c_2\left[y_{1}\left(\frac{\partial^{4} y_{2}}{\partial x_{0}^{4}} + 4\frac{\partial^{4} y_{1}}{\partial x_{0}^{3} \partial x_{1}}\right) + y_{2} \frac{\partial^{4} y_{1}}{\partial x_{0}^{4}} + \frac{\partial y_{1}}{\partial x_{0}}\left(\frac{\partial^{3} y_{2}}{\partial x_{0}^{3}} + 3\frac{\partial^{3} y_{1}}{\partial x_{0}^{2}\partial x_{1}}\right) + \frac{\partial^{3} y_{1}}{\partial x_{0}^{3}}\left(\frac{\partial y_{2}}{\partial x_{0}} + \frac{\partial y_{1}}{\partial x_{1}}\right)\right] \nonumber \\
&+& \frac{1}{2}a_1 y_{1}^{2}\frac{\partial y_{1}}{\partial x_{0}} + b_3\left[\frac{1}{2} y_{1}^{2}\frac{\partial^{2} y_{1}}{\partial x_{0}^{2}} + y_{1}\left(\frac{\partial y_{1}}{\partial x_{0}}\right)^{2}\right]  + c_2\left[\frac{1}{2} y_{1}^{2}\frac{\partial^{4}\eta_{1}}{\partial x_{0}^{4}} + y_{1} \frac{\partial y_{1}}{\partial x_{0}} \frac{\partial^{3} y_{1}}{\partial x_{0}^{3}}\right]. \label{nons}
\end{eqnarray}
We now group terms of the same power in $\zeta$ in eq. (\ref{e22}), and set each of them to zero. To the lowest order (i.e. $\zeta^0$) we obtain:
\begin{equation}
L_{0} y_{1} = 0,
\label{e23}
\end{equation}
which, from the form of $L_{0}$ given in formula (\ref{lin}), suggests a solution:
\begin{equation}
y_{1} = \phi(x_{1},...,x_{N},t_{1},...,t_{N})\,exp(i\theta) + c.c.
\label{e24}
\end{equation}
To the order $\zeta^{2}$ we obtain:
\begin{equation}
 L_{0} y_{2} + L_{1} y_{1} = - N_{1}.
 \label{e25}
\end{equation}
This solution can also be expressed as:
\begin{equation}
L_{0} y_{2} = -i\left[\frac{\partial D(\omega_{r},k)}{\partial \omega_{r}} \frac{\partial \phi}{\partial t_{1}} - \frac{\partial D(\omega_{r},k)}{\partial k} \frac{\partial \phi}{\partial x_{1}}\right] e^{i\theta} - \Omega \phi^{2}e^{2i\theta} \hskip 0.1truecm + c.c,
\label{e26}
\end{equation}
where:
\begin{eqnarray}
\frac{\partial D(\omega_r,k)}{\partial \omega_r} &=& -i, \nonumber \\ \frac{\partial D(\omega_r,k)}{\partial k} &=& ia_1 - 2b_1k + 4c_1k^{3}, \nonumber \\
\Omega &=& ia_2 - 2b_2k^{2} + 2c_2k^{4}.
\label{e27}
\end{eqnarray}
A non-secularity condition implies setting the coefficient of $e^{i \theta}$ in (\ref{e26}) to zero, i.e.:
\begin{equation}
 -i\left[\frac{\partial D(\omega_r,k)}{\partial \omega_r} \frac{\partial \phi}{\partial t_{1}} - \frac{\partial D(\omega_{r},k)}{\partial k} \frac{\partial \phi}{\partial x_{1}}\right] = 0.
 \label{e28}
\end{equation}
This leads to:
\begin{equation}
 \frac{\partial \phi}{\partial t_{1}} +(a_1 + 2b_1ik - 4c_1 ik^{3}) \frac{\partial \phi}{\partial x_{1}} = 0.
 \label{e29}
\end{equation}
In terms of the group velocity $V_g$, we can also express the last relation as:
\begin{equation}
 \frac{\partial \phi}{\partial t_{1}} + V_{g}\frac{\partial \phi}{\partial x_{1}} = 0,
 \label{e30}
\end{equation}
where $V_{g} = \frac{d w}{d k}  =  -\frac{\partial D_{k}}{\partial D_{w_{r}}}$. Eq. (\ref{e30}) admits a uniform solution for $y_{2}$ given by:
 \begin{equation}
y_{2} = - \frac{\Omega \phi^{2}e^{2i\theta}}{D(2w_{r},2k)} + F(x_{1},...,x_{N},t_{1},...,t_{N})e^{i \theta} + G(x_{1},...,x_{N},t_{1},...,t_{N}) \hskip 0.1truecm + c.c,
\label{e31}
 \end{equation}
where F and G are complex and real functions respectively of higher-order scales, which can be obtained from higher-order perturbations, and:
\begin{equation}
D(2\omega_{r},2k) =  -2i\omega_{r} + 2ia_{1}k - 4b_{1}k^{2} + 16c_{1}k^{4}.
\label{e32}
\end{equation}
To the order $\zeta^{3}$ we have:
\begin{equation}
 L_{0} y_{3}+L_{1} y_{2} + L_{2} y_{1} = - N_{2}.
 \label{e33}
\end{equation}
Substituting $y_{1}$ given by (\ref{e24}), and $y_{2}$ given by (\ref{e31}), in eq. (\ref{e33}), we find the second non-secularity condition:
\begin{eqnarray}
 &i&\left[\frac{\partial D(w_{r},k)}{\partial w_{r}} \frac{\partial \phi}{\partial t_{2}} - \frac{\partial D(w_{r},k)}{\partial k} \frac{\partial \phi}{\partial x_{2}}\right]
+(b_1 - 6c_1k^{2}) \frac{\partial^{2} \phi}{\partial x_{1}^{2}} + i\left[\frac{\partial D(w_{r},k)}{\partial w_{r}} \frac{\partial F}{\partial t_{1}} - \frac{\partial D(w_{r},k)}{\partial k} \frac{\partial F}{\partial x_{1}}\right] 
\nonumber \\ &=& -\left[a_1ik + 3(c_2k^{4} - b_3k^{2}) + (9b_2k^{2} - 3a_2ik - 43c_2k^{4})\frac{\Omega}{D(2w_{r},2k)} \right] \left| \phi\right|^{2} \phi - \Sigma G \phi, \nonumber \\
\label{e34}
\end{eqnarray}
where $\Sigma = a_2ik  - b_2k^{2} + c_2k^{4}$. Since the linear operators $L_i$ and the nonlinear functions $N_i$ are composed of first-order spatial and temporal derivatives, as well as derivatives of higher orders, the relation governing the variation of the slow mode $G$ can be obtained from eq. (\ref{e33}) and is given by:
\begin{equation}
\frac{\partial G}{\partial t_{1}} + a_{1} \frac{\partial G}{\partial x_{1}} = 0.
 \label{e35}
\end{equation}
We assume that both $F$ and $G$ depend on higher scales through $\phi$, for we are interested in the case with first-order pertubation\cite{kawa1}. considering that $\phi$, $F$ and $G$ 
depend on $x_{1}$ and $t_{1}$ through $\eta_{1} \equiv x_{1} - v t_{1}$, the first non-secularity condition (\ref{e28}) is always satisfied such that the third equation on the left-hand side of eq. (\ref{e34}) vanishes. Furthermore, solving eq. (\ref{e35}) gives:
\begin{equation}
G (\eta_{1}, x_{2},...,x_{N},t_{2},...,t_{N}) = \gamma (x_{2},...,x_{N},t_{2},...,t_{N}),
\label{e36}
\end{equation}
provided that the group and phase velocities are differents. Replacing eq. (\ref{e36}) and eq. (\ref{e28}) in the second non-secularity condition (\ref{e34}), we obtain a nonlinear equation for the complex amplitude $\phi$:
\begin{eqnarray}
 &i&\left[\frac{\partial D(w_{r},k)}{\partial w_{r}} \frac{\partial \phi}{\partial t_{2}} - \frac{\partial D(w_{r},k)}{\partial k} \frac{\partial \phi}{\partial x_{2}}\right]
+(A_{2} - 6A_{3}k^{2}) \frac{\partial^{2} \phi}{\partial x_{1}^{2}}  
\nonumber \\ &=& -\left[a_1ik + 3(c_2k^{4} - b_3k^{2}) + (9b_2k^{2} - 3a_2ik - 43c_2k^{4})\frac{\Omega}{D(2w_{r},2k)} \right] \left| \phi\right|^{2} \phi - \Sigma \gamma \phi. \nonumber \\
\label{e37}
\end{eqnarray}
Let us introduce two new coordinates namely $\tau = t_{2} - \vartheta x_{2}$ and $ \eta = \zeta x $, with these eq. (\ref{e37}) becomes:
\begin{equation}
 i\frac{\partial \phi}{\partial \tau} + (\Gamma_{1} - i \Gamma_{2}) \frac{\partial^{2} \phi}{\partial \eta^{2}}  + (\beta_{1} + i\beta_{2}) \left| \phi\right|^2\phi + (\Sigma_{1} + i\Sigma_{2}) \gamma \phi = 0,
\label{e42}
\end{equation}
with:
\begin{eqnarray}
\Gamma_{1} &=& \frac{\lambda_{1} (b_1 - 6c_1k^{2})}{\lambda_{1}^{2} + \lambda_{2}^{2}}, \hskip 0.25truecm \Gamma_{2} = \frac{ \lambda_{2} (b_1 - 6c_1k^{2})}{\lambda_{1}^{2} + \lambda_{2}^{2}}, \\
\lambda_{1} &=& 1 + \vartheta (4 c_1k^{3} - 2b_1 k^{2}), \hskip 0.25truecm \lambda_{2} = \vartheta a_1, \\
\beta_{1} &=&  \frac{\lambda_{1} \upsilon_{1} - \lambda_{2} \upsilon_{2}}{\lambda_{1}^{2} + \lambda_{2}^{2}}, \hskip 0.25truecm \beta_{2} = \frac{\lambda_{1} \upsilon_{2} + \lambda_{2} \upsilon_{1}}{\lambda_{1}^{2} + \lambda_{2}^{2}}, \\
\upsilon_{1} &=&  3(c_2k^{4} - b_3k^{2}) + \left[\frac{104 b_2c_2 k^{6} - a_2^{2}k^{2} - 18 b_2^{2} k^{4} - 86 c_2^{2} k^{8}}{16c_1k^{4} - 4b_1k^{2}}\right], \\
\upsilon_{2} &=&   a_1k + \left[\frac{15 a_2b_2k^{3} - 49 b_2 c_2k^{5}}{16a_{1} k^{4} - 4b_1k^{2}}\right], \\
\Sigma_{1} &=& \frac{\lambda_{1}(c_2k^{4} - b_2k^{2}) - \lambda_{2} (a_2k)}{\lambda_{1}^{2} + \lambda_{2}^{2}}, \hskip 0.25truecm \Sigma_{2} = \frac{\lambda_{1} (a_2k) + \lambda_{2} (c_2k^{4} - b_2k^{2})}{\lambda_{1}^{2} + \lambda_{2}^{2}}. 
\end{eqnarray}
Eq. (\ref{e42}), which resembles a complex-coefficient nonlinear Schr\"odinger equation with damping or growth \cite{stenf1,kivsh1}, is more exactly a dissipative cubic complex Ginzburg-Landau (DCCGL) equation~\cite{cgl1,kur,cross,pism,bohr,kram,chat1}.\\ 
Although the DCCGL eq. (\ref{e42}) stands for the general nonlinear equation derived from the multiple-scale expansion and the reductive-perturbation expansion applied on the partial-differential equation (\ref{e14}), not all terms in this equation are relevant to the specific problem under study. Namely the quantity $\gamma$ in eq. (\ref{e42}) is actually an arbitrary constant, i.e. does not depends on any of the coefficients of eq. (\ref{e14}). In fact $\gamma$ was introduced in eq. (\ref{e36}) to take into account the contribution of $G$ in the multiple-scale expansion solution. This contribution can readily be ingored within the framework of the multiple-scale expansion theory, such that the system dynamics is governed by the CCGL equation:
\begin{equation}
 i\frac{\partial \phi}{\partial \tau} + (\Gamma_{1} - i \Gamma_{2}) \frac{\partial^{2} \phi}{\partial \eta^{2}}  + (\beta_{1} + i\beta_{2}) \left| \phi\right|^2\phi = 0.
\label{e42a}
\end{equation}
The CCGL eq. (\ref{e42a}) admits exact nonlinear periodic solutions of the general form \cite{vel}: 
\begin{equation}
 \varphi (\xi) = \phi(\xi) e^{i\theta(\xi,\tau)} , \hskip 0.3truecm \xi = \eta - C \tau,
 \label{e87a}
\end{equation}
where $\phi(\xi)$ is the wave envelope and $\theta(\xi,\tau)$ is its carrier. Substituting (\ref{e87a}) in eq. (\ref{e42a}), separating real from imaginary parts and integrating the two resulting nonlinear equations, we obtain:
\begin{equation}
 \phi(\xi) = \sqrt{\Phi(\xi)}, \hskip 0.2truecm \theta(\xi,\tau) = \beta_0 ln(\Phi(\xi)) + A, \label{dna}
\end{equation}
where:
\begin{eqnarray}
 \Phi(\xi) &=& \frac{6 \beta_0 (\Gamma_{1}^{2} + \Gamma_{2}^{2})}{\beta_{2}\Gamma_{1} + \beta_{1}\Gamma_{2}} [(e'_{1} - e'_{3}) dn^{2}(\sqrt{e'_{1} - e'_{3}} \xi, \kappa_0)] \nonumber \\
&+& \frac{2 B \Gamma_{2} - C \Gamma_{1} - 36 e'_{1} A_{1} (\Gamma_{1}^{2} + \Gamma_{2}^{2})}{4 (\beta_{2}\Gamma_{1} + \beta_{1}\Gamma_{2})}. \label{dnab}
\end{eqnarray}
We can rearrange the periodic solution (\ref{e87a}) with $\phi$ and $\theta$ given as in (\ref{dna}), to find:
\begin{equation}
 \varphi(\xi) = \Phi(\xi)^{\frac{1}{2} + i \beta_0} e^{i A}. \label{varph0}
\end{equation}
$\kappa_0$ in the above relations is defined as:
\begin{equation}
\kappa_0 = \sqrt{\frac{e'_{2} - e'_{3}}{e'_{1} - e'_{3}}}, 
\end{equation}
where $e'_{1,2,2}$ are three real roots of a cubic polynomial \cite{vel}. Note also the presence of $A_1$ and $A$ in the periodic solution (\ref{dna}), the parameter $A$ is an arbitrary chirp (and hence can be discarded) while $\beta_0$ is obtained in terms of the model parameters as:
\begin{equation}
\beta_0 = -M\pm\sqrt{\frac{1}{2}+M^2}, \hskip 0.2truecm M = \frac{3}{4}\frac{\beta_1\Gamma_1-\beta_2\Gamma_2}{\beta_1\Gamma_2+\beta_2\Gamma_1}. \label{eqana}
\end{equation}
With the exact periodic solution of the CCGL eq. (\ref{e42a}), we obtain $y(x,t)$ (i.e. the solution to eq. (\ref{e14})) by constructing the series (\ref{sumap}) keeping only the relevant power in $\zeta$ (i.e. $\zeta^2$). Using (\ref{e24}) and (\ref{e31}) together with the expression of $\varphi$ given in (\ref{varph0}), we find:
\begin{equation}
 y(x,t) = \zeta \Phi(x,t)^{\frac{1}{2} + i \beta_0} - \frac{\zeta^2\Omega}{D(2\omega_r, 2k)}\,\Phi(x,t)^{1 + 2i \beta_0} + c.c. \label{finsol}
 \end{equation}
The analytical solution (ref{finsol}) contains several parameters, all dependent on characteristic paramaters of the model. It is therefore not possible, within the framework of the present work, to discuss features of the solution considering all these paramaters. Nevertheless, to have a sight of some typical profiles of the solution, we considered varying the paramater $\beta_0$ which, according to formula (\ref{eqana}), can take on positive and negative values depending on magnitudes of $\beta_1$, $\beta_2$, $\Gamma_1$ and $\Gamma_2$. Also the periodic function $dn$ is known to tend to "sech" as the modulus $\kappa_0$ tends to one. We have therefore focused on varying $\beta_0$ for two different values of $\kappa_0$ and fixing all other parameters in (\ref{finsol}) to arbitrary values, just for the sake of illustration since several combinations of values of these parameters give rise to a rich varieties of shape profiles, all of which cannot be displayed here. \\  
Figs. \ref{fig:three} and \ref{fig:four} display shape profiles of the real part of $y(\xi)$ considering four different negative values of $\beta_0$, for $\kappa_0=0.9$ (fig. \ref{fig:three}) and $\kappa_0=0.99$ (fig. \ref{fig:four}). On the other hand, figs. \ref{fig:five} and \ref{fig:six} are shape profiles of the real part of $y(\xi)$ for positive values of $\beta_0$ and same values of $\kappa_0$ as in the two previous figures. 
As the figures suggest, the modulation is more and more pronounced as $\beta_0$ increases. For small values of this parameter profiles of the real part of $y(\xi)$ are cnoidal waves, including cnoidal-wave solutions to the Korteweg-de Vries equation for certain values of parameters in (\ref{finsol}). 
\section{\label{sec:level4}Conclusion}
Instability of a thin liquid film falling down a heated substrate has been among the recent attractive topics in fluid mechanics from standpoints of fundamental theory and experiment. From the standpoint of theory it is an open-flow hydrodynamic problem which is convectively unstable~\cite{chang,greg,cras,kaba}, leading to a sequence of wave transitions beginning with the amplification of small-amplitude noises at the inlet, the filtering of linear stability and ultimately a secondary modulational instability that transforms the primary wave field into a solitary-wave. The system dynamics has been discussed at length in the recent past within the framework of the so-called Benney equation, based on the long-wave instability picture introduced by Kapitza and Kapitza~\cite{kap2} and studied in detail by Benjamin~\cite{benj,benj1}. \\
According to the long-wave instability picture the dynamics of the thin liquid film over a heat substrate will depend mainly on the Marangoni number, the capillary effects, the Reynolds number (related to gravity) and surface tensions. When the plate is horizontal the Reynolds number is zero, the system dynamics in this case is governed by the competition between the Marangoni effect and surface tensions. \\
In this work we considered the last problem in the specific context when the Biot number is very small and the plate is uniformly heated. We established that in this regime, the Benney equation can admit two distinct families of periodic nonlinear wave structures. The first one is a train of topological solitary waves similar to cnoidal waves of the Korteveg-de Vries equation, and the second is a train of envelope solitons with spatio-temporal modulations. The last type of nonlinear wave structure was obtaianed via the multiple-scale exapansion combined with the reductive perturbation method, which led to a cubic complex Ginzburg-Landau equation. This equation is familiar in fluid mechanics, where it describes the dynamics of several distinct structures in fluids including modulated solitary-wave envelopes, spiral-wave structures and so on.\\
In our study the temperature along the horizontal substrate was assumed homogeneous. However, in some applications it can be useful to consider both mass and heat transfers along the substrate. To this last point, it is well known \cite{orong1,orong4,thiele1,legros2} that the temperature variation along the substrate will induce thermocapillary convections in addition to capillary effects related to surface tensions. Despite numerical evidences \cite{orong1,orong4,thiele1} that thermocapillary instabilities do not prevent the formation of fingering ripples and their buildups into rivulet structures, changes in characteristic parameters of solitary waves, as well as nonlinear periodic structures similar to those proposed in this study, are to be expected. This later problem, as well as the issue of thin-liquid film flow on a slightly inclined (i.e. the case with nonzero Reynolds number) and inhomogeneously heated substrate, are some interesting open problems deserving a theoretical attention.\\
To end, we would like to stress that our main objective in this study was to investigate analytical solutions to the Benney equation for the thin-liquid film flow problem on a uniformly heated horizontal surface. A large number of previous studies have considered this equation numerically, and several distinct shape profiles are proposed in the literature from numerical simulations assuming various flow configurations. The present work therefore aims at proposing analytical solutions to be compared with the numerous recent numerical results on the Benney equation. 

\begin{acknowledgments}
The work of A. M. Dikand\'e is supported by the Alexander von Humboldt (AvH) foundation. The authors wish to thank T. C. Kofan\'e for enriching exchanges. The Laboratory of Research on Advanced Materials and Nonlinear Sciences (LaRAMaNS), is partially funded by The World Academy of Sciences (TWAS), Trieste Italy.
\end{acknowledgments}

\end{document}